\documentclass[graybox]{svmult}

\usepackage{amssymb}
\usepackage[T1]{fontenc}
\usepackage[utf8]{inputenc}
\usepackage[intlimits]{amsmath}
\usepackage{enumerate}
\usepackage[all]{xy}

\usepackage{mathptmx}       
\usepackage{helvet}         
\usepackage{courier}        
\usepackage{type1cm}        
\usepackage{makeidx}         
\usepackage{graphicx}        
\usepackage{multicol}        
\usepackage[bottom]{footmisc}

\makeindex             

\begin{document}

\title*{Factorization method and general second order linear difference equation}

\author{Alina Dobrogowska and  Mahouton Norbert Hounkonnou}

\institute{Alina Dobrogowska \at Institute of Mathematics, University of Białystok, Ciołkowskiego 1M, 15-245 Białystok, Poland, \email{alina.dobrogowska@uwb.edu.pl}
\and Mahouton Norbert Hounkonnou \at International Chair in Mathematical Physics and Applications (ICMPA--UNESCO Chair), University of Abomey-Calavi, 072 BP 50, Cotonou, Republic of Benin, \email{norbert.hounkonnou@cipma.uac.bj}}

\maketitle

\abstract{ This paper addresses an investigation on a factorization method for difference equations. It is proved that some classes of second order linear difference operators, acting in Hilbert spaces,  can be factorized using a pair of  mutually adjoint first order difference operators. These classes encompass equations of hypergeometic type describing classical orthogonal polynomials of a discrete variable.}

%

\section{Introduction}

The description of many problems in physics and mathematics, especially in probability, gives rise to difference equations. Difference equations relate to differential equations as discrete mathematics relates to continuous mathematics.
The study of differential equations shows that  even supposedly elementary examples can be hard to solve. By contrast, elementary difference equations are relatively easy to deal with. In general, the  interest in difference equations can be justified for a number of reasons. Difference equations frequently arise when modelling  real life situations. Since difference equations are readily handled by numerical methods, a standard approach to solving a nasty differential equation is to convert it to an approximately equivalent difference equation. 

A peculiar question in the field of differential or difference equations then remains to find appropriate analytical methods for their exact solvability.
For differential or difference equations having polynomial solutions,
it is well known that their solvability is closely related to the factorizability of their associated operators, (see \cite{bang0} and references therein).

In the last few decades it was given a  more
prominent place in the discussion of operator factorization methods for solving second order differential 
 or 
difference equations, the concept of which goes back to 
Darboux \cite{Dar}. Later the method was rediscovered many times, in particular by the founders of quantum mechanics, (see Dirac \cite{Dir}, Schr\"odinger \cite{Schr}), while solving the Schrödinger equation to study the angular momentum or the harmonic oscillator. In the work \cite{InHu}, which is now considered to be fundamental, Infeld and Hull  summarized the quantum mechanical applications of the method. Later this technique was extended, see \cite{ Mielnik-2, MieNie, Fer}. Some results were obtained also for  $q$-difference and more general difference equations  \cite{b1, b12, bang1, bang2, B4, B5, DoJa, GolOdz, Mil-1, Mil-2}.
In addition, special cases such as the factorization of Jacobi operators were also investigated \cite{GeTe}.
If the operator in a second order linear
ordinary differential or difference equation can be factorized, the problem of solving the equation
is reduced to solving two first order linear equations; the latter can readily be
solved.

Therefore, a nodal point in the application of this method  consists in the existence
of a pair of first order differential or difference operators,  which   the second order differential or difference operator decomposes into as their product, (see (\ref{2}) in this work).

Using this method, we are here able to  find the explicit solutions  (\ref{3ee}), via (\ref{3e}),  to the eigenvalue problem (\ref{3f}) in a simple way. 
 For additional readings,  see monographs \cite{Dong-1, 16, Mielnik-3}.

This work is an extension of a previous work \cite{DoJa}. Some results obtained in \cite{b1, B14A, B14, GolOdz} are  used, and adapted to our context. 

The paper  is organized as follows. In Section 2,  a detailed investigation of the factorization method applied to  second order difference operators is given. In Section 3,  our main results are described. Under given assumptions, the problem of  operator  factorization is solved.

\section{Basic tools}

 In this section, in the beginning, we introduce some notations and recall some basic facts about the  factorization method.
Let $\ell_k(\mathbb{Z},\mathbb{R})$  and $\ell_k(\mathbb{Z},\mathbb{C})$, $k\in \mathbb{N}\cup\{0\}$, be the  sets of real-valued  and complex-valued sequences $\{x(n)\}_{n\in\mathbb{Z}}$, respectively. We define the scalar product on  $\ell_k(\mathbb{Z},\mathbb{C})$  as follows:
\begin{equation}
\label{0}
\left<x | y\right>_k:=\sum_{n=a}^{b} \overline{x(n)}y(n)\rho_k(n),
\end{equation}
where $a,b\in\mathbb{Z},\; (a<b),$ and $\rho_k$  is a weight function. We assume that  the weight 
sequence satisfies  the Pearson difference equation
\begin{equation}
\label{0a}
\bigtriangleup \left(b_k(n)\rho_k(n)\right)=\left( c_k(n)-b_k(n)\right) \rho_k(n),
\end{equation}
and   the recursion relation
\begin{equation}
\label{0b}
\rho_{k-1}(n)=c_k(n)\rho_k(n),
\end{equation}
where $\{b_k\}$  and $\{c_k\}$ are some  real-valued sequences.
Moreover, the function $\rho_k$  fulfills the boundary conditions
\begin{equation}
\label{0c}
b_k(a)\rho_k(a)=b_k(b+1)\rho_k(b+1)=0.
\end{equation}
The forward and backward difference operators are defined by
 \begin{align}
 & \bigtriangleup  x(n):=\left({\bf S^{+}}-{\bf{1}}\right) x(n)=x(n+1)-x(n),\\
 &  \bigtriangledown x(n):=\left({\bf{1}}-{\bf S^{-}}\right) x(n)=x(n)-x(n-1),
 \end{align}
where the shift operators 
 \begin{equation}
 {\bf S^{\pm}}x(n):=x(n\pm 1).
 \end{equation}
 
We  want to apply the  factorization method to  the second order difference operators ${\bf H_k}: \ell_k(\mathbb{Z},\mathbb{C})\longrightarrow \ell_k(\mathbb{Z},\mathbb{C})$ given by
\begin{equation}
\label{1}
{\bf H_k}:=z_k(n){\bf S^{+}}+w_k(n){\bf S^{-}}+v_k(n),
\end{equation}
where $\{z_k\},\{w_k\}$  and $\{v_k\}$ are  real-valued sequences,  $k\in \mathbb{N}\cup\{0\}$.
Introducing  the annihilation  ${\bf A_k}: \ell_k(\mathbb{Z},\mathbb{C})\longrightarrow \ell_{k-1}(\mathbb{Z},\mathbb{C}),$ and creation operators  ${\bf A^*_k}: \ell_{k-1}(\mathbb{Z},\mathbb{C})\longrightarrow \ell_k(\mathbb{Z},\mathbb{C})$ (also called lowering and raising operators,  respectively), 
we rewrite the above operators ${\bf H_k}$ in the form
\begin{equation}
\label{2}
{\bf H_k}:={\bf A_k^*}{\bf A_k}+\alpha_k={\bf A_{k+1}}{\bf A_{k+1}^{*}}+\alpha_{k+1},
\end{equation}
where $\alpha_k$ are  real constants.
We  construct  the annihilation  operator as
\begin{align}
&\label{3} {\bf A_k}:=\bigtriangleup +f_k(n)={\bf S^{+}}+f_k(n)-1,
\end{align} 
where $\{f_k\}\in \ell_k(\mathbb{Z},\mathbb{R})$.

We seek the adjoint operator ${\bf A^*_k}$ of  ${\bf A_k}$,  obeying:
\begin{equation}
\label{3a}
\left< {\bf A_k^*} x_{k-1}| y_{k}\right>_k=\left<  x_{k-1}| {\bf A_k} y_{k}\right>_{k-1}.
\end{equation}
A simple computation using (\ref{3a}) yields
\begin{eqnarray}
\left<  x_{k-1}| {\bf A_k} y_{k}\right>_{k-1}&=&
\sum_{n=a}^{b} \overline {x_{k-1}(n)} y_k(n+1)\rho_{k-1}(n)\cr
&+&\sum_{n=a}^{b} \overline {x_{k-1}(n)} (f_k(n)-1)y_k(n)\rho_{k-1}(n)\cr
&=&
\sum_{n=a+1}^{b+1} b_k(n)\overline{x_{k-1}(n-1)} y_k(n)\rho_{k}(n)\cr &+&\sum_{n=a}^{b} (f_k(n)-1)c_k(n)\overline{x_{k-1}(n)}y_k(n)\rho_{k}(n)\cr
&=&
\left< \left(b_k(n){\bf S^{-}}+\left(f_k(n)-1\right)c_k(n) \right)  x_{k-1}| y_{k}\right>_k ,
\end{eqnarray}
where we applied the formulas   (\ref{0a})-(\ref{0c}). Finally, we obtain the explicit expression for the adjoint operator (also called creation operator)
\begin{eqnarray}
\label{3b} {\bf A_k^{*}} &  =&-b_k(n)\bigtriangledown +b_k(n)+\left(f_k(n)-1\right)c_k(n)\cr
 &  =&b_k(n){\bf S^{-}}+\left(f_k(n)-1\right)c_k(n).
\end{eqnarray}
This type of factorization was presented in detail in papers \cite{GolOdz}, \cite{B4}, \cite{B14A} for $\tau$--, $q$-- and $(q,h)$--cases, respectively. Moreover,  different cases, when the sequence $b_k$ does not depend on parameter $k$, were considered  in \cite{DoJa}.

 The operator ${\bf H_k}={\bf A_k^*}{\bf A_k}+\alpha_k$ is selfadjoint  on $\ell_k(\mathbb{Z},\mathbb{C}).$  Its eigenvalue equation reads:
\begin{equation}
\label{3f}
{\bf H_k} x^l_k(n)=\lambda^l_k x^l_k(n).
\end{equation}
It is well known that the factorization gives us the eigenfunctions and   corresponding eigenvalues.  Indeed,   the eigenvalue problem for the chain of operators (\ref{2}) is equivalent to the two following equations:
\begin{align}
&\label{3c} {\bf A_k^*}{\bf A_k} x^l_k(n)= \left(\lambda^l_k -\alpha_k \right) x^l_k(n),\\
&\label{3d} {\bf A_{k+1}}{\bf A_{k+1}^{*}} x^l_k(n)=  \left(\lambda^l_k -\alpha_{k+1} \right) x^l_k(n).
\end{align}
Solving the first order homogeneous linear equation
\begin{equation}
\label{3e}
{\bf A_k} x^0_k(n)= 0,
\end{equation}
we observe that (\ref{3c}), (\ref{3d}) and (\ref{3e}) imply that the functions
\begin{equation}
\label{3ee}
 x^{k-p}_k(n)= {\bf A_{k}^{*}} {\bf A_{k-1}^{*}}\dots {\bf A_{p+1}^{*}}x^0_p(n)
\end{equation}
are solutions of  the eigenvalue problem (\ref{3f}) for  the eigenvalues $\lambda_k^{k-p}=\alpha_p$.

\section{Factorization of operators}

In this section,   we solve the factorization problem (\ref{2}) under some assumptions.  Finding a general solution remains a cumbersome task.

Comparing  the coefficients of ${\bf S^{-}}$, ${\bf S^{+}}$ and ${\bf 1}$  on both sides of the expression (\ref{2}), we  obtain the necessary and sufficient conditions  for the existence of  a factorizing pair of first order difference operators, $({\bf A_k^*},{\bf A_k}),$  as follows:
\begin{eqnarray}
\label{5} f_{k+1}(n)-1&=&\frac{b_k(n)}{b_{k+1}(n)}\left(f_k(n-1)-1\right),\\
\label{6} c_{k+1}(n)&=&\frac{b_{k+1}(n)}{b_{k}(n)}c_k(n-1),\\
 \label{7} b_k(n) \!-\! b_{k+1}(n+1)\!&=&\!\alpha_{k+1} \!-\!  \alpha_{k}\!+\!\frac{b_k(n)}{b_{k+1}(n)}\left(\!f_k(n-1)-1\!\right)^2 \!c_k(n-1)\!\nonumber\\
&-&\!\left(\!f_k(n)-1\!\right)^2 \!c_k(n).
\end{eqnarray}
The conditions (\ref{5}) and (\ref{6}) give us  the transformation formulas for the sequences $\{f_k\}$ and $\{c_k\}$  as below:
\begin{equation}
	f_k(n)= \prod_{i=1}^{k}\frac{b_{k-i}(n-i+1)}{b_{k-i+1}(n-i+1)}\Big(f_0(n-k)-1\Big) +1
	\end{equation}
	and
	\begin{equation}
	c_k(n) = \prod_{i=1}^{k}\frac{b_{k-i+1}(n-i+1)}{b_{k-i}(n-i+1)}c_0(n-k).
	\end{equation}

\subsection{Example 1}
We assume that $b_{k+1}(n)=b_{k}(n)=:b_0(n)$, i.e.  the  sequence $\{b_k\}$ does not depend on parameters $k$, 
see \cite{DoJa}.
We  show that, under this assumption, we  can find a general solution to the factorization problem 
(\ref{2}), i.e. we can solve the conditions  (\ref{5})-(\ref{7}).

We have:
	\begin{equation}
	\left \{
	\begin{array}{l}
	\displaystyle
f_{k+1}(n) = f_k(n-1)\\
	c_{k+1}(n) = c_k(n-1)
	\end{array}
	\right.
	\end{equation}
yielding
		\begin{equation}
	\left \{
	\begin{array}{l}
	\displaystyle
	f_{k}(n) = f_0(n-k)\\
	c_{k}(n) = c_0(n-k)	.
	\end{array}
	\right.
	\end{equation}
Now, let us   solve the third condition. 
The requirement (\ref{7}),  using the substitution
$G_k(n)=\left(f_k(n)-1\right)^2 c_k(n)- b_{0}(n+1),$
is equivalent to the equation
\begin{equation}
G_k(n)=G_k(n-1)+\alpha_{k+1}-\alpha_{k}.
\end{equation}
By iterating we find
\begin{equation}
G_k(n)=G_k(0)+n\left( \alpha_{k+1}-\alpha_{k} \right).
\end{equation}
This gives us a formula for the sequence  $\{b_0\}:$

\begin{equation}
\label{e1}
b(n+1)=\left(f_0(n-k)-1\right)^2 c_0(n-k)-G_k(0)-n\left( \alpha_{k+1}-\alpha_{k} \right).
\end{equation}
But the left-hand side of the expression (\ref{e1}) does not depend on the parameter $k.$ Then, we obtain the following sequence of  conditions on the sequences $\{f_0\}$ and $\{c_0\}:$
\begin{eqnarray}
\label{8a}
\left(f_0(n)-1\right)^2 c_0(n)-G_0(0)-n\left( \alpha_{1}-\alpha_{0} \right)&=&
\left(f_0(n-k)-1\right)^2 c_0(n-k)\cr
&-&G_k(0)-n\left( \alpha_{k+1}-\alpha_{k} \right), 
\end{eqnarray}
for all $k\in\mathbb{N}$.
By introducing $F(n)=\left(f_0(n)-1\right)^2 c_0(n),$ we can write the above equation in the form:
\begin{equation}
F(n)=F(n-k)+G_0(0)-G_k(0)-n\left(\alpha_{k+1}-  \alpha_{k}-\alpha_1+\alpha_0\right).
\end{equation}
For  $k=1$, we get 
\begin{equation}
F(n)=F(n-1)+G_0(0)-G_1(0)-n\left(\alpha_{2}-  2\alpha_{1}+\alpha_0\right).
\end{equation}
Next, by iteration we find
\begin{equation}
\label{9a}
F(n)=F(0)+n\left( G_0(0)-G_1(0) \right) -\frac{n(n+1)}{2}  \left(\alpha_{2}-  2\alpha_{1}+\alpha_0\right).
\end{equation}
We then arrive at a relationship between the sequences $\{f_0\}$ and $\{c_0\}:$
\begin{equation}
\label{9aa}
\left( f_0(n)-1 \right)^2 c_0(n)=F(0)+n\left( G_0(0)-G_1(0) \right) -\frac{n(n+1)}{2}  \left(\alpha_{2}-  2\alpha_{1}+\alpha_0\right).
\end{equation}

In addition, substituting (\ref{9a}) to (\ref{8a}) (because it is valid for all $k\in\mathbb{N}$), we find a recurrence relation on  the constants $\alpha_k:$ 
\begin{equation}
\alpha_{k+1}=\alpha_k+\alpha_1-\alpha_0 +k \left(\alpha_2-2\alpha_1+\alpha_0\right)
\end{equation}
and the form of the constant 
\begin{equation}
G_k(0)=G_0(0)+k\left(G_1(0)-G_0(0)\right)-\frac{k(k-1)}{2} \left(\alpha_2-2\alpha_1+\alpha_0\right).
\end{equation}
A straightforward calculation  affords
\begin{equation}
\alpha_k=\alpha_0+k\left(\alpha_1-\alpha_0\right)+\frac{k(k-1)}{2} \left(\alpha_2-2\alpha_1+\alpha_0\right).
\end{equation}

 To sum up,  the construction presented in (\ref{2}) provides the chain of operators ${\bf H_k}$ parametrized by the freely chosen sequence $\{c_0\}$ and real parameters $\alpha_0$, $\alpha_1$, $\alpha_2$, 
$F(0)$, $G_0(0)$ and $G_1(0)$.

\subsection{Example 2}
Let us consider a case when $f_k(n)\equiv 0.$ Then, the conditions (\ref{5})-(\ref{7}) can be rewritten  in the form:
\begin{align}
&\label{5a} b_{k+1}(n)=b_k(n)=:b_0(n),\\
&\label{6a} c_{k}(n)=c_0(n-k),\\
&\label{7a}  b_0(n) - b_0(n+1)=\alpha_{k+1}-  \alpha_{k}+c_0(n-k-1)-c_0(n-k).
\end{align}
This is a special case of Example $1$.
We find $b_0(n)$ by induction in the following form:
	\begin{equation}
	 b_{0}(n) = b_0(0) - n(\alpha_{k+1} - \alpha_k) - c_0(-1-k) + c_0(n-1-k).
	\end{equation}
From (\ref{9aa}) we obtain that the sequence $\{c_0\}$ is a polynomial of degree two: 
\begin{equation}
c_0(n)=c_0(0)+n\left( G_0(0)-G_1(0) \right)-\frac{n(n+1)}{2}\left(\alpha_2-2\alpha_1+\alpha_0\right).
\end{equation}

	Then, the relation ${\bf H_k}x^l_k(n) = \lambda^l_kx^l_k(n)$ is equivalent to
	\begin{eqnarray}
	\Big(-b_0(n)\nabla + b_0(n) - c_0(n-k)\Big)\Delta x^l_k(n) = (\lambda^l_k - \alpha_k)x^l_k(n),
	\end{eqnarray}	
and the eigenvalue problem (\ref{3f}) is reduced to the difference equation of hypergeometric type:
\begin{equation}\label{hypeqn}
-b_0(n)  \bigtriangledown \bigtriangleup x^l_k(n)-\left(c_0(n-k)-b_0(n)\right)\bigtriangleup  x^l_k(n)+
\left(\alpha_k-\lambda^l_k\right)  x^l_k(n)=0.
\end{equation}
It is not difficult to see that 
 $b_0$  is a second degree polynomial  while the  difference 
 $c_0(n-k)-b_0(n)$ is a first degree polynomial.
From equation (\ref{3e}), we find that the ground state is a constant sequence  $\{x_k^0(n)\equiv 1\}$ (normalized to one) with  the sequence of eigenvalues $\{\lambda_k^0=\alpha_k=\alpha_0+k\left(\alpha_1-\alpha_0\right)+\frac{k(k-1)}{2} \left(\alpha_2-2\alpha_1+\alpha_0\right)\}$. Expression (\ref{3ee}) gives us a formula for polynomials
\begin{equation}
P_l(n)=x_k^l(n)=\prod_{i=0}^{l-1}\left( b_0(n){\bf S^-}-c_0(n-k+i)\right)1
\end{equation}
corresponding to eigenvalues $\lambda_k^l=\alpha_{k-l}.$
Using the identity 
$ \bigtriangledown \bigtriangleup = \bigtriangleup  \bigtriangledown$ we transform the above equation (\ref{hypeqn}) into the standard form
\begin{equation}
\sigma (n) \bigtriangleup  \bigtriangledown  x^l_k(n)+\tau (n)  \bigtriangleup x^l_k(n) +\lambda x^l_k(n)=0,
\end{equation}
where
\begin{align}
\sigma (n) & =-b_0(n) =  \frac 12 \left(\alpha_2-2\alpha_1+\alpha_0\right) n^2+n\left(G_1(0)-G_0(0) +\alpha_1-\alpha_0 \right. \nonumber\\
& \left.- \frac 12 \left(\alpha_2-2\alpha_1+\alpha_0\right)\right)-b_0(0),\\
 \tau (n) &  =b_0(n)-c_0(n-k)=  \left(\alpha_0-\alpha_1+(1-k)\left(\alpha_2-2\alpha_1+\alpha_0\right)\right) n \nonumber\\ 
& +b_0(0)-c_0(0)+k\left(G_0(0)-G_1(0)\right)+\frac{k(k-1)}{2}\left(\alpha_2-2\alpha_1+\alpha_0\right)  , \\
\lambda & = \alpha_k-\lambda^l_k=\alpha_k-\alpha_{k-l}= l\left(\alpha_1-\alpha_0\right) \nonumber\\
& + \left( kl-\frac{l(l+1)}{2} \right) \left(\alpha_2-2\alpha_1+\alpha_0\right)
=-l\left( \tau'(n)+\frac{l-1}{2}\sigma''(n)\right).
\end{align}
It is well known that the above equation (\ref{hypeqn}) describes  classical orthogonal polynomials of a discrete variable such as the Charlier, Meixner, Kravchuk, Hahn polynomials.  See  \cite{NiSu, b12} for more details.

\subsection{Example 3}
We assume that $b_{k+1}(n):=\gamma_k b_k(n)$, where $\gamma_k$ is some constant different from zero and one.
Then, we get:
	\begin{equation}\label{l}
	f_{k}(n) - 1 = \prod_{i=1}^{k}\gamma^{-1}_{k-i}(f_0(n-k)-1),
	\end{equation}
	\begin{equation}\label{m}
	c_k(n) = \prod_{i=1}^{k}\gamma_{k-i}c_0(n-k),
	\end{equation}
and
	\begin{eqnarray}\label{n}
		b_k(n) - \gamma_k b_{k}(n+1) &= & \alpha_{k+1} - \alpha_k + \gamma^{-1}_k(f_k(n-1)-1)^2c_k(n-1)\cr &-&(f_k(n)-1)^2c_k(n).
	\end{eqnarray}
Using previous results, (see Example $1$), we  solve by analogy the above difference equation. 
By introducing $R_k(n)=\left( f_k(n)-1\right)^2c_k(n) -\gamma_k b_k(n+1)$, we can write this equation in the form
\begin{equation}
R_k(n)=\gamma_k^{-1} R_k(n-1)+\alpha_{k+1}-\alpha_k.
\end{equation}
By  iterating we find
\begin{equation}
R_k(n)=\gamma_k^{-n}R_k(0)+\frac{1-\gamma_k^{-n}}{1-\gamma_k^{-1}}\left(\alpha_{k+1}-\alpha_k\right).
\end{equation}
From here, expressing everything by the initial data we obtain
\begin{eqnarray}
b_k(n+1)&=&\gamma_k^{-1}\gamma_{k-1}^{-1}\dots \gamma_{0}^{-1} \left( f_0(n-k)-1\right)^2c_0(n-k)-\gamma_k^{-n-1} R_k(0)\cr
&-&\gamma_k^{-1} \frac{1-\gamma_k^{-n}}{1-\gamma_k^{-1}}\left(\alpha_{k+1}-\alpha_k\right).
\end{eqnarray}
This must be consistent with the initial assumption, i.e. $b_{k+1}(n):=\gamma_k b_k(n).$
Then, we get the condition on the sequences $\{f_0\}$ and $\{c_0\}:$
\begin{eqnarray}
&&\gamma_{k+1}^{-1}\gamma_{k}^{-1}\dots \gamma_{0}^{-1} \left( f_0(n-k-1)-1\right)^2c_0(n-k-1)
-\gamma_{k+1}^{-n-1} R_{k+1}(0)\cr
&-&\gamma_{k+1}^{-1} \frac{1-\gamma_{k+1}^{-n}}{1-\gamma_{k+1}^{-1}}\left(\alpha_{k+2}-\alpha_{k+1}\right)
=
\gamma_{k-1}^{-1}\dots \gamma_{0}^{-1} \left( f_0(n-k)-1\right)^2c_0(n-k)-\gamma_k^{-n} R_k(0)\cr
&-&\frac{1-\gamma_k^{-n}}{1-\gamma_k^{-1}}\left(\alpha_{k+1}-\alpha_k\right).
\end{eqnarray}
Again, by entering the following auxiliary function:
$$
S_k(n)= \gamma_{k-1}^{-1}\dots \gamma_{0}^{-1} \left( 
f_0(n-k)-1\right)^2c_0(n-k),$$ we get  a recursion relation for $S_k:$
\begin{eqnarray}
S_k(n)&=&\gamma_{k+1}^{-1}\gamma_{k}^{-1} S_k(n-1) -\gamma_{k+1}^{-n-1} R_{k+1}(0)+\gamma_k^{-n} R_k(0)\cr
&-&\gamma_{k+1}^{-1} \frac{1-\gamma_{k+1}^{-n}}{1-\gamma_{k+1}^{-1}}\left(\alpha_{k+2}-\alpha_{k+1}\right)
+\frac{1-\gamma_k^{-n}}{1-\gamma_k^{-1}}\left(\alpha_{k+1}-\alpha_k\right).
\end{eqnarray}
By iteration,
\begin{eqnarray}
S_k(n)&=&\left( \gamma_{k+1}^{-1}\gamma_{k}^{-1}\right)^n S_k(0)
-\gamma_{k+1}^{-n-1} \frac{1-\gamma_k^{-n}}{1-\gamma_k^{-1}} R_{k+1}(0)
+\gamma_k^{-n} \frac{1-\gamma_{k+1}^{-n}}{1-\gamma_{k+1}^{-1}} R_k(0) \cr
&-&\gamma_{k+1}^{-1} \left(\alpha_{k+2}-\alpha_{k+1}\right) 
\sum_{i=0}^{n-1} \left( \gamma_{k+1}^{-1}\gamma_{k}^{-1}\right)^i \frac{1-\gamma_{k+1}^{-n+i}}{1-\gamma_{k+1}^{-1}}\cr
&+&\left(\alpha_{k+1}-\alpha_k\right)
\sum_{i=0}^{n-1} \left( \gamma_{k+1}^{-1}\gamma_{k}^{-1}\right)^i \frac{1-\gamma_k^{-n+i}}{1-\gamma_k^{-1}}.
\end{eqnarray}

\section{Concluding remarks}
In this work, we have   investigated  a factorization method for difference equations, adapting and extending previous  results known in the literature. We have showed  that some classes of second order linear difference operators, acting in Hilbert spaces,  are factorizable using a pair of  mutually adjoint first order difference operators. These classes encompass equations of hypergeometic type describing classical orthogonal polynomials of a discrete variable. Other classes of difference equations are still under consideration, and will be in the core of our forthcoming papers.

An interesting outlook on which we  are also working is the extension of this  scheme  to  classes of higher order
difference equations. It is in particular  expected that this method for fourth order  equations may allow  to derive what one can call Krall–Laguerre–Hahn polynomials.

\section*{Acknowledgements}
 AD is partially supported by the Santander Universidades grant. She also would like to thank the organizers of  ICDDEA 2017 in Amadora, Portugal, for their hospitality.

\end{document}